\documentclass[prl,twocolumn,showpacs,amssymb,floats]{revtex4}

\usepackage{epsf}
\usepackage{amsmath}

\def \be#1\ee {\begin{equation}#1\end{equation}}
\def \bea#1\eea {\begin{eqnarray}#1\end{eqnarray}}
\newcommand{\corr}[1]{\langle #1\rangle}
\newcommand{\Corr}[1]{\left<#1\right>}

\newcommand{\Tr}{\mathop{\rm Tr}}
\newcommand{\tr}{\mathop{\rm tr}}

\newcommand{\vp}{\varphi}

\begin{document}
\bibliographystyle{prsty}

\title{Quantum correction to the Kubo formula in closed mesoscopic systems}

\author{Mikhail A. Skvortsov}

\affiliation{L. D. Landau Institute for Theoretical Physics,
Moscow 117940, Russia}

\begin{abstract}
We study the energy dissipation rate in a mesoscopic system
described by the parametrically-driven random-matrix Hamiltonian $H[\vp(t)]$
for the case of linear bias $\vp=vt$.
Evolution of the field
$\vp(t)$ causes interlevel transitions leading to energy
pumping, and also smears the discrete spectrum of the Hamiltonian.
For sufficiently fast
perturbation this smearing exceeds the mean level spacing and the dissipation
rate is given by the Kubo formula. We calculate the quantum correction
to the Kubo result that reveals the original discreteness of the energy
spectrum.
The first correction to the system viscosity scales $\propto v^{-2/3}$
in the orthogonal case and vanishes in the unitary case.
\end{abstract}


\maketitle

The Kubo formula~\cite{Kubo} is one of the cornerstones
of modern condensed matter physics.
It is a standard tool for calculating various linear response
functions, with conductivity as a prototypical example.
Based on the lowest-order perturbation theory for a continuous spectrum,
Kubo approach connects the kinetic response of a system with respect
to some external field $\vp(t)$ to the equilibrium correlator
of generalized currents.

Application of the Kubo formula essentially relies on the
assumption of a continuous spectrum~\cite{Mahan}.
(For the case of a discrete spectrum it gives either zero,
when the frequency of the external field $\vp(t)$ is out of resonance
with any pair of energy levels, or infinity otherwise,
indicating breakdown of the linear response theory.)
The widespread usage of the Kubo formula for macroscopic objects
is justified by the smallness of the mean level spacing
$\Delta\propto\hbar^d/L^d$ ($d$ is the dimensionality
of space and $L$ is the system size) which is usually smaller
than the inelastic width $\Gamma_{\rm in}$ of energy levels.
In an electron system, the inelastic smearing is due to
interaction
as well as escape to reservoirs operative for open systems.

For closed microscopic systems (e.g., quantum dots),
the condition of continuous spectrum
is violated at sufficiently low temperatures when the interaction-induced
smearing $\Gamma_{\rm in}$ becomes smaller than $\Delta$.
In this case individual discrete levels are well resolved and the Kubo
formula may fail to describe the dissipation in the system.
Such a situation had recently been discussed in the context
of vortex dynamics in layered superconductors at low
temperatures~\cite{FS97,LO98,KL99,SIB02}.

In a closed system with $\Gamma_{\rm in}\ll\Delta$, the mechanism of
dissipation depends on the rate $v=d\vp/dt$ of variation of the external
field $\vp(t)$. For sufficiently slow perturbations with $v \ll v_K$
(the velocity $v_K$ depends on the sensitivity of the
spectrum to the change of $\vp$ and will be defined below),
the system adiabatically follows the spectrum of its instantaneous
Hamiltonian $H[\vp(t)]$, and the dissipation is due to rare Landau--Zener
transitions between individual levels~\cite{LZ}.
Fast perturbations with $v\gg v_K$ cause multiple transitions
between energy levels thereby transforming the discrete spectrum of
the stationary Hamiltonian into a featureless quasi-continuous spectrum,
where the dissipation can be obtained with the help of the Kubo formula.

\begin{figure}[b]
\centerline{\epsfbox{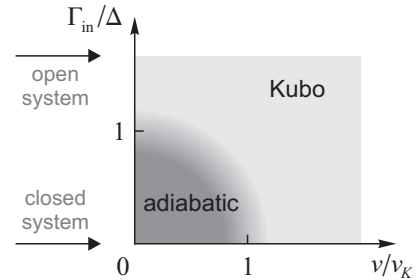}}
\caption{Position of the adiabatic and Kubo regimes
as a function of the inelastic width $\Gamma_{\rm in}$ and velocity $v$.}
\label{F:diagram}
\end{figure}

These two regimes had been studied by Wilkinson~\cite{Wilkinson88}
who calculated the dissipation in the adiabatic ($v\ll v_K$) and
Kubo ($v\gg v_K$) regimes assuming the Hamiltonian from the Wigner-Dyson
random-matrix ensembles~\cite{Mehta}.
In the Kubo regime, the energy dissipation rate is given
by the linear-response formula:
\be
  W_{\rm Kubo} = \frac\beta2 \pi C(0)v^2,
\label{W-Kubo}
\ee
where $\beta=1$ for the orthogonal (GOE)
and $2$ for the unitary (GUE) ensembles,
and $C(0) \equiv \Corr{(\partial E_{i}/\partial \vp)^2}/\Delta^2$
is the variance of the level velocity normalized by the mean level
spacing $\Delta$.
(The system of units with $\hbar=1$ is used throughout the Letter.)
Equation (\ref{W-Kubo}) suggests that $C(0)$ has
the meaning of the generalized conductance~\cite{Simons-Altshuler}.
It determines the critical velocity $v_K \sim \Delta/\sqrt{C(0)}$
separating the adiabatic and Kubo regimes of dissipation.
In the adiabatic limit,
the dissipation rate nontrivially depends on the symmetry
of the Hamiltonian~\cite{Wilkinson88}:
\be
  W_\beta = c_\beta v^{1+\beta/2},
\label{W-Zener}
\ee
where $c_1=(\pi/4)\Gamma(\frac34)[C(0)]^{3/4}\Delta^{1/2}$
and $c_2=\pi C(0)$.
Hence the dissipation is superohmic for the GOE, while for the GUE
it remains ohmic, exactly coinciding with $W_{\rm Kubo}$ despite
a very different mechanism of dissipation.

A schematic diagram indicating the regions of the adiabatic and Kubo
regimes as a function of the inelastic width $\Gamma_{\rm in}$ and
velocity $v$ of external perturbation is shown in Fig.~\ref{F:diagram}.
Counterintuitively, the linear-response Kubo formula does not describe
the low-velocity behavior of closed systems at $\Gamma_{\rm in}\ll\Delta$!

In this Letter we address the question of how does the
discreteness of energy levels of a stationary ($v=0$) Hamiltonian
manifest itself in the Kubo regime ($v\gg v_K$) when the levels are
smeared into a quasi-continuous spectrum.
We show that the relative correction to the high-velocity
asymptotics~(\ref{W-Kubo}) can be regularly expanded
in integer powers of $(v_K/v)^{2/3}$.
For the orthogonal symmetry, the first (one-loop) quantum correction
to $W_{\rm Kubo}$ is given by
\be
  \frac{W_1}{W_{\rm Kubo}} =
  1 + \frac{\Gamma(\frac13)}{3^{2/3}}
    \left( \frac{v_K}{v} \right)^{2/3} + \dots ,
\label{W/W}
\ee
where the omitted terms have the order of $O[(v_K/v)^{4/3}]$,
and the crossover velocity is defined as
$v_K \equiv (2^{1/2}/\pi^2) \Delta/\sqrt{C(0)}$.
Thus, the remaining correlations in the quasi-continuous spectrum
enhance dissipation at $v\gg v_K$, with a gradual crossover to the
superohmic regime (\ref{W-Zener}) at $v\ll v_K$.
In the unitary case, the first and the second (two-loop)
quantum correction to the Kubo result (\ref{W-Kubo}) vanish,
making it tempting to conjecture that the identity
$W_2\equiv W_{\rm Kubo}$ holds for all velocities.

Quantum corrections to the quasiclassical properties of disordered
systems had been the subject of intense studies in the last
decades~\cite{WL,Efetov}.
In treating these phenomena, the nonlinear $\sigma$-model
was proven to be the most powerful tool both in the perturbative and
nonperturbative regimes (where it often is the only possible approach).
Three versions of the $\sigma$-model based on the supersymmetry~\cite{Efetov},
replica~\cite{replicas}, and Keldysh~\cite{Horbach-Schoen,Kamenev-Andreev}
techniques had been proposed for noninteracting systems.

The problem of energy pumping by the parametrically-driven
Hamiltonian $H[\vp(t)]$ belongs to the class of non\-equi\-lib\-rium problems,
that dictates the choice of the Keldysh formalism as a solution tool.
We will derive the Keldysh $\sigma$-model for the
parametrically-driven random-matrix Hamiltonian
and show that its saddle-point solution
yields the kinetic equation for the distribution
function, reproducing the Kubo result (\ref{W-Kubo}).
Fluctuations near this saddle point are responsible for the quantum
correction to the Kubo formula leading to Eq.~(\ref{W/W}).

We consider a time-dependent matrix Hamiltonian
\be
  H[\vp(t)] = H_1 \cos k\vp(t) + H_2 \sin k\vp(t),
\label{Ham}
\ee
where $H_1$ and $H_2$ are the $N\times N$ matrices from the same GOE
($H^T=H$) distributed with the probability density
$P(V) \propto \exp[-(\pi^2/2N\Delta^2) \tr V^2]$,
and $k = (\pi\sigma/\Delta) (2/N)^{1/2}$.
The density of states for an instant Hamiltonian
is given by the Wigner semicircle:
$\rho(E) = \Delta^{-1} [1-\pi^2E^2/4N^2\Delta^2]^{1/2}$,
with $\Delta$ being the mean level spacing at the center of the band.
The dispersion of the matrix elements
\be
  \biggl<\left|\frac{\partial H_{ij}}{\partial\vp}\right|^2\biggr> =
  \sigma^2 (1+\delta_{ij})
\label{sigma}
\ee
is independent of the matrix size $N$.
The generalized conductance is then $C(0)=2\sigma^2/\Delta^2$.
The states of the Hamiltonian are filled by $N/2$ noninteracting
fermions.

The time evolution governed by $\vp(t)$ will change the
state of the system and eventually pump some energy into it.
The energy of the system will grow unless the inelastic interaction
with the thermal-bath is taken into account.
This interaction will establish a nonequilibrium steady state.
Remarkably, however, that the energy dissipation rate is
independent of the resulting distribution and hence can be
calculated as a mean growth rate of the total system energy
of noninteracting fermions~\cite{Wilkinson88}.

Within the Keldysh formalism, the system is described by the action
\be
  S[\Psi] = -i \int_{-\infty}^\infty dt \;
    \Psi^\dagger(t)
    \left[ i\tau_3\frac{\partial}{\partial t} - H[\vp(t)] \right]
      \sigma_3 \Psi(t) ,
\label{S[psi]}
\ee
where $\Psi(t)$ is a Grassmanian $4N$-vector field acting in the direct
product of the index space of the Hamiltonian, Keldysh (K) space and
particle-hole (PH) space. The Pauli matrices in the K and PH spaces
are denoted by $\sigma_i$ and $\tau_i$, respectively. The Keldysh
forward-backward structure is a necessary ingredient of the dynamic formalism
ensuring the correct normalization $Z \equiv \int D\Psi e^{-S} = 1$
in the absence of quantum sources,
whereas the PH grading is introduced in order to handle
the orthogonal symmetry of the Hamiltonian~\cite{Efetov}.

Derivation of the $\sigma$-model, which is a low-energy effective theory
for the action~(\ref{S[psi]}), is a standard procedure.
One has to average $Z$ over the Hamiltonian~(\ref{Ham}), decouple it by the
$4\times4$ matrix $Q_{tt'}$ and integrate over fermions $\Psi$.
Keeping the terms which are finite in the limit $N\to\infty$
and assuming linear bias $\vp=vt$
we arrive at the following action (the weight is
$e^{-S}$) for the $\sigma$-model \cite{sigma-comment}:
\be
  S = \frac{\pi i}{2\Delta} \Tr \hat E \tau_3 Q
    + \frac{\pi\Omega^3}{8\Delta} \int dt\, dt'\, (t-t')^2
    \tr Q_{tt'} Q_{t't} ,
\label{sigma-model}
\ee
where $\Omega^3 \equiv \pi\sigma^2v^2/\Delta = (\pi/2) C(0) v^2 \Delta$.
The first term in Eq.~(\ref{sigma-model}) is the standard random-matrix
action~\cite{AK00} which is responsible for the whole spectral
statistics. The second term is of kinetic origin; it accounts
for interlevel transitions of the time-dependent Hamiltonian $H[\vp(t)]$.
The field theory with the action~(\ref{sigma-model})
describes both the adibatic and Kubo regimes
of dissipation on an equal footing.
It is controlled by the single parameter
\be
  \frac{\Omega}{\Delta} = \frac1\pi \left(\frac{v}{v_K} \right)^{2/3}
\ee
which will be used hereafter as a measure of velocity $v$.

In the stationary case ($\Omega=0$), the Keldysh Green function $Q$
is diagonal in the energy representation~\cite{Kamenev-Andreev}:
\be
  \Lambda =
    \left( \begin{array}{cc}
      1 & 2F \\
      0 & -1
    \end{array} \right)
    \otimes \tau_3 ,
\label{Lambda}
\ee
where $F(E)=1-2f(E)$, and $f(E)$ is the electron distribution function.
The evolution of the distribution function at $\Omega\neq0$ is described
by the saddle point of the action~(\ref{sigma-model}).
Varying with respect the constraint $Q^2=\openone$
one obtains the saddle-point equation $[Q,\delta S/\delta Q]=0$.
Seeking the solution in the form (\ref{Lambda}),
we obtain the equation for the
distribution function $f_{tt'}$:
\be
\label{spe}
  \left( \frac{\partial}{\partial t} + \frac{\partial}{\partial t'} \right)
  f_{tt'}
  =
  - \Omega^3 (t-t')^2 f_{tt'} .
\ee
The same equation had been obtained in Ref.~\onlinecite{Kravtsov01a}
starting with the diagrammatic technique.
Performing the Wigner transformation
$f(E,t) = \int d\tau e^{iE\tau} f_{t+\tau/2,t-\tau/2}$
we arrive at the kinetic equation
\be
  \frac{\partial f(E,t)}{\partial t}
  = \Omega^3 \frac{\partial^2f(E,t)}{\partial E^2} .
\label{diffusion}
\ee
This is a diffusion equation in the energy space, with the
bare ``diffusion coefficient'' ${\cal D}_0=\Omega^3$.
The rate of energy pumping for the system described by
the kinetic equation~(\ref{diffusion}) is given by
\be
  W = \int E \frac{\partial f(E,t)}{\partial t} \frac{dE}{\Delta}
    = \frac{\Omega^3}{\Delta} ,
\label{W-MF}
\ee
that coincides with the result of the Kubo formula (\ref{W-Kubo}).

The kinetic equation~(\ref{diffusion}) is a true saddle-point of
the action for all velocities $v$. The answer (\ref{W-MF}) for the
dissipation rate predicted is valid, however, only in the Kubo regime,
and is completely wrong in the adiabatic regime. The reason is
that the saddle-point approximation is
justified by the large value of the parameter $\Omega/\Delta$.
In the adiabatic regime, the saddle-point approximation fails
and one has to take all the $Q$-manifold into account. As a result
of this procedure, the answer (\ref{W-Zener}) should be reproduced.
Note the interchange of steps with respect to the Wilkinson's
derivation~\cite{Wilkinson88}: He first calculates the probability
of the Landau-Zener transition and then averages
it over the distribution of avoided crossings. Here we, instead, first
average over randomness and then extract the dissipation rate from the
field theory (\ref{sigma-model}).
Thus it is a challenging problem to demonstrate how the
adiabatic result (\ref{W-Zener}) should be obtained from the
field theory (\ref{sigma-model}).

Quantum correction to the Kubo result (\ref{W-MF}) in the limit
$\Omega/\Delta\gg1$ ($\Delta/\Omega$ is the loop expansion parameter)
can be obtained in the regular way by expanding
over Gaussian fluctuations near the saddle point.
The matrix $Q$ can be written
as~\cite{Kamenev-Andreev}
\be
  Q = U_F^{-1} P U_F,
  \qquad
  U_F =
  \left( \begin{array}{cc}
    1 & F \\ 0 & -1
  \end{array} \right) ,
\ee
with the Hermitian matrix $P$ having an additional symmetry
$P^T=\sigma_1\tau_2P\tau_2\sigma_1$ imposed by the orthogonal
symmetry of the Hamiltonian.
We choose the standard rational parameterization of the $P$-matrix,
$P=\sigma_3\tau_3 (1+V/2)(1-V/2)^{-1}$, with the unit Jacobian.
The matrix $V$ is explicitly given by
\be
  V = \left( \begin{array}{cccc}
    0 & a & b & 0 \\
    -a^\dagger & 0 & 0 & -b^T \\
    -b^\dagger & 0 & 0 & a^T \\
    0 & b^* & -a^* & 0
  \end{array} \right) ,
\ee
where the inner (outer) grading corresponds to the PH (K) space.
The elements $a_{tt'}$ and $b_{tt'}$ describe cooperon and diffuson
modes, respectively. Their bare propagators
read~\cite{Vavilov99,Kravtsov01a,Kravtsov01b,bb-comment}
\begin{align}
  \corr{a_{t_1t_2}a^*_{t_3t_4}} &=
    \frac{2\Delta}{\pi}
    \, \delta(t_{13}+t_{24}) \, \theta(t_{13}) \,
    e^{-\Omega^3 ({t_{13}^3}/{3} + t_{13} t_{14}^2)} ,
\label{cooperon}
\\
  \corr{b_{t_1t_2}b^*_{t_3t_4}} &=
    \frac{2\Delta}{\pi}
    \, \delta(t_{13}-t_{24}) \, \theta(t_{13}) \,
    e^{-\Omega^3 t_{13} t_{12}^2} ,
\label{diffuson}
\end{align}
where $t_{ij} \equiv t_i-t_j$.
In the stationary case ($\Omega=0$),
Eqs.~(\ref{cooperon}) and (\ref{diffuson}) describe
the standard cooperon and diffuson in the time domain. The exponential
decay of the correlators at the time scale $\Omega^{-1}$ is a manifestation
of dephasing by the time-dependent perturbation~\cite{Vavilov99,Kravtsov01a}.

\begin{figure}
\epsfxsize=0.6\hsize
\centerline{\epsfbox{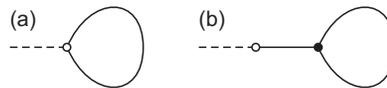}}
\caption{1-loop corrections to the system energy.
The solid lines stand for the propagators (\ref{cooperon})
and (\ref{diffuson}),
the dashed line denotes for the source field $\kappa(t)$,
and the open and solid vertexes come from the terms $S_\kappa$
and $S$, respectively.}
\label{F:1loop}
\end{figure}

\begin{figure}
\epsfxsize=0.6\hsize
\centerline{\epsfbox{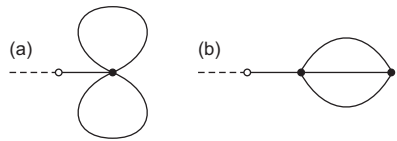}}
\caption{2-loop corrections to the system energy in the unitary case.}
\label{F:2loops}
\end{figure}

The system energy $\corr{E}$ (apart from an additive constant)
can then be obtained as a functional derivative with respect
to the quantum source field $\kappa(t)$:
\be
\label{<E>}
  \corr{E(t)}
  = \left. \frac12 \frac{\delta Z[\kappa]}{\delta\kappa(t)} \right|_{\kappa=0} ,
  \qquad
  Z[\kappa] = \int e^{-S-S_\kappa} DQ ,
\ee
where the source action
$S_\kappa = (\pi/2\Delta) \Tr \kappa \hat E \sigma_1 \tau_3 Q$.
Indeed, the presence of the Pauli matrix $\sigma_1$ cuts
the off-diagonal in the Keldysh space elements of the matrix $Q$,
thus projecting onto the upper right (Keldysh) block containing the
distribution function renormalized by quantum fluctuations
(the lower left block vanishes due to causality).

At the saddle point, Eq.~(\ref{<E>}) reproduces the result (\ref{W-MF}).
The one-loop diagrams are shown in Fig.~\ref{F:1loop}.
The diagram (a) coming from the expansion of the sourse term $S_\kappa$
contains either $\corr{a(t_1,t_2)a^*(t_1,t_3)}$
or $\corr{b(t_1,t_2)b^*(t_1,t_3)}$ which are proportional to $\theta(0)$.
In the Keldysh formalism, the Heaviside $\theta$-function of zero
argument evaluates to zero, that is related with the causality
of the theory~\cite{AK00}.
Calculating the diagram (b) we obtain for the quantum correction
to the dissipation rate:
\be
  \frac{\delta W_1}{W_{\rm Kubo}} =
    \frac{\Gamma(\frac13)}{3^{2/3}\pi}
    \frac{\Delta}{\Omega} ,
\label{total-rational}
\ee
which after rewriting in terms of velocity leads to Eq.~(\ref{W/W}).
This result may be interpreted in terms of renormalization
of the bare ``diffusion coefficient'' in Eq.~(\ref{diffusion}):
${\cal D}_0 \to {\cal D}_0(1+\delta W_1/W_{\rm Kubo})$,
bearing a natural analogy with the weak localization
phenomena. In our case, the ratio $\Omega/\Delta$ plays the role
of the dimensionless conductance
as it controls the expansion in terms of the diffusive modes.

In the unitary case, the diagram of Fig.~\ref{F:1loop}(b) vanishes
indicating the absence of the one-loop quantum correction
to the Kubo result. In the two-loop approximation,
only the diagrams shown in Fig.~\ref{F:2loops} contribute
to the dissipation rate for the GUE.
Each of them has the order of $(\Delta/\Omega)^2$,
but their sum is zero. Thus, for the case of the unitary
symmetry, the two-loop correction also vanishes.
Taking into account the coincidence between the low-
and high-velocity asymptotics (\ref{W-Kubo}) and (\ref{W-Zener})
for $\beta=2$, such a cancelation is a strong argument in favor
of the exact identity $W_2 = W_{\rm Kubo}$ valid for
all velocities.
At present we cannot prove this conjecture but we hope
that this can be done by an accurate analysis of the $\sigma$-model
(\ref{sigma-model}).
We conjecture that the independence of the dissipation
on the velocity $v$ might be another manifestation of
the peculiar properties of the unitary group~\cite{Zirnbauer}.
It is worth mentioning that the direct quantum-mechanical
calculation of the transition rates at $v\sim v_K$ {\em prior}
to disorder averaging seems completely impossible.
Therefore, the identity $W_2=W_{\rm Kubo}$ for the averaged
dissipation rate would be a highly nontrivial fact.

The results obtained may be relevant for the description of heating
effect in closed quantum dots whose shape is being changed by the
low-frequency external perturbation. Vortex motion in impure
superconductors is another field of application, where the conjecture
$W_2=W_{\rm Kubo}$ would indicate the independence of the dissipative
flux-flow conductivity on the velocity of vortex motion~\cite{FS97,SIB02}.

Summarizing, we have developed the Keldysh $\sigma$-model approach
to study energy pumping in the para\-met\-ric\-ally-driven
random-matrix ensembles.
We calculated the leading quantum correction to the high-velocity
dissipation, which reveals the original discreteness of the spectrum
of the stationary Hamiltonian. This correction emerges in the one-loop
approximation for the GOE, and is absent within the two-loop accuracy
for the GUE.

I am grateful to D.~M.~Basko, Ya.~M.~Blanter, D.~A.~Ivanov, M.~V.~Feigel'man,
V.~E.~Kravtsov, A.~I.~Larkin, and Yu.~V.~Nazarov for illuminating discussions.
Financial support from
the SCOPES program of Switzerland, the Dutch
Organization for Fundamental Research (NWO), the Russian
Foundation for Basic Research under grants 01-02-17759
and 02-02-06238, the program
``Quantum Macrophysics'' of the Russian Academy of Sciences, the
Russian Ministry of Science, and the Russian Science Support Foundation
is acknowledged.


\begin{thebibliography}{100}
\bibitem{sigma-comment}
The general form of the kinetic term valid for an arbitrary dependence
of $\vp(t)$ is $S_{\rm kin} = (\beta\pi^2 C(0)/16)
\int dt\, dt'\, [\vp(t)-\vp(t')]^2 \tr Q_{tt'} Q_{t't}$.
In this form it is valid both for the GOE and GUE.
\bibitem{bb-comment}
There is also a finite average $\corr{b_{t_1t_2}b_{t_3t_4}}$ but it is
irrelevant for the calculation of the mean dissipation rate.
\end{thebibliography}
\end{document}